\documentclass[pra,twocolumn,showpacs,preprintnumbers,amsmath,amssymb]{revtex4}

\usepackage{graphicx}
\usepackage{dcolumn}
\usepackage{bm}
\usepackage{color}

\def\Fbox#1{\vskip1ex\hbox to 8.5cm{\hfil\fboxsep0.3cm\fbox{%
\parbox{8.0cm}{#1}}\hfil}\vskip1ex\noindent}  
\newcommand{\eq}[1]{(\ref{#1})}
\newcommand{\Eq}[1]{Eq.~(\ref{#1})}
\newcommand{\Fig}[1]{Fig.~\ref{#1}}
\newcommand{\Figs}[1]{Figs.~\ref{#1}}
\newcommand{\Ref}[1]{Ref.~\cite{#1}}
\newcommand{\Refs}[1]{Refs.~\cite{#1}}

\def\beq{\begin{equation}} \def\eeq{\end{equation}}
\def\bea{\begin{eqnarray}} \def\eea{\end{eqnarray}}
\def\bse{\begin{subequations}} \def\ese{\end{subequations}}
\def\||{\parallel}
\let\p\partial

\def\<{\left\langle} \def\>{\right\rangle}
\def\({\left(} \def\){\right)}
\def\[{\left[} \def\]{\right]}

\makeatletter
\AtBeginDocument{\@ifpackageloaded{natbib}{\ifNAT@numbers\if@filesw\immediate\write\@auxout{\string\global\string\NAT@numberstrue}\fi\fi}{}}
\makeatother
\begin{document}

\title{Instability Crossover of Helical Shear-flow in Segregated Bose-Einstein Condensates}

\author{Shinsuke Hayashi$^1$, Makoto Tsubota$^2$, and Hiromitsu Takeuchi$^1$}
\email[]{hirotake@sci.osaka-cu.ac.jp}
\affiliation{$^1$Department of Physics, Osaka City University, Sumiyoshi-ku, Osaka 558-8585, Japan\\
$^2$Department of Physics and The Osaka City University Advanced Research Institute for Natural Science and Technology (OCARINA), Osaka City University, Sumiyoshi-ku, Osaka 558-8585, Japan}%
\date{\today}

\begin{abstract}
We theoretically study the instability of helical shear flows,
 in which one fluid component flows along the vortex core of the other, in phase-separated two-component Bose-Einstein condensates at zero temperature.
 The helical shear flows are hydrodynamically classified into two regimes:
(1)   a helical vortex sheet, where the vorticity is localized on the cylindrical interface and the stability is described by an effective theory for ripple modes, and (2)
a core-flow vortex with the vorticity distributed in the vicinity of the vortex core,
 where the instability phenomena are dominated only by the vortex-characteristic modes (Kelvin and varicose modes).
 The helical shear-flow instability shows  remarkable competition among different types of instabilities in the crossover regime between the two regimes.
\end{abstract}


\maketitle

\section{Introduction}
Kelvin-Helmholtz instability (KHI), one of the most fundamental instabilities in fluid dynamics,
 occurs in the presence of shear flow between two immiscible fluids \cite{cKHI1}.
 When the relative velocity between the two fluids is sufficiently large,
 a vortex sheet existing along the interface between the two becomes unstable and develops into characteristic roll-up patterns of eddies.
 Because of the universal applicability of fluid dynamics,
  KHI may occur not only in classical fluids but also in quantum fluids such as superfluids \cite{helium,2010TakeuchiPRB}.
 However, the KHI in superfluids (namely, quantum KHI) can be distinct from that in a classical fluid
 since the dynamics is governed by  macroscopic quantum effects, superfluidity, and vortex quantization.
 Recently, quantum KHI has been proposed in phase-separated two-component Bose-Einstein condensates (BECs)
 consisting of two distinguishable Bose particles \cite{2010TakeuchiPRB,2010SuzukiPRA}.
 Quantum KHI is realized in the presence of a relative superfluid velocity
 between the two components.
 When the relative velocity exceeds a critical value,
 a flat interface between the two superfluids changes into  characteristic sawtooth waves without energy dissipation,
 leading to the formation of single-quantized vortices from the peaks and troughs of the waves,
 which is quite different from the nonlinear development in KHI in classical fluids.

Although quantum KHI is a quantum counterpart of KHI in classical fluids,
 exotic hydrodynamic instabilities without classical counterparts also offer  interesting possibilities  \cite{phys_re}.
 Such instabilities can occur as a direct result of the quantum effects in superfluids: superfluidity and vortex quantization.
 For example, superfluidity (the disappearance of friction) makes it possible to realize a counterflow of two miscible fluid components,
 called countersuperflow, in two-component BECs.
 The instability of countersuperflow leads to exotic instability phenomena \cite{CSI1,CSI2} and
 has been realized very recently in two-component BECs \cite{Hamner2011PRL}.
 Another possibility can be realized in the presence of quantized vortices.
 Kelvin wave (or Donnelly-Glaberson) instability \cite{Glaberson,D_book},
 which causes a deformation of a vortex from a straight line into a helix, is one such fundamental  type of instability discussed in different superfluids \cite{araki, FinneRepProgr2006, 2009Takeuchi}.

\begin{figure}[htbp]
 \centering
 \includegraphics[width=0.45 \linewidth]{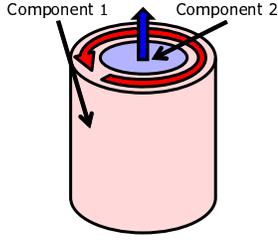}
 \caption{
 (Color online) Schematic diagram of helical shear flow in phase-separated two-component BECs.
 The first component forms a quantized vortex and the second component flows along the vortex core.
}
 \label{fig:schematic}
\end{figure}

 In this work, we theoretically study the instability of  helical shear flow in phase-separated two-component BECs,
 in which one component flows along the core of a quantized vortex of the other component forming a cylindrical interface,
 as is illustrated in Fig. \ref{fig:schematic}.
 This state is called helical shear flow since the trajectory along the local relative velocity between the two fluids is helical.
 Helical shear flow is a flow state stabilized by the inviscid flows and  vortex quantization in superfluids.
 Helical shear flows may naturally arise in turbulent or nonequilibrium states in segregated condensates
 with a large population imbalance between the two components (e.g., see \Refs{VortexformationJLTP, VortonPRA, TachyonPRL, TachyonJLTP}).
 A helical shear flow becomes dynamically unstable when the relative velocity along the vortex core exceeds a critical value.
 The instability phenomena change drastically depending on the rotational velocity and the linear velocity of the first and second components, respectively.
 In this paper, we develop   the phase diagram and investigate the nonlinear development of the helical shear-flow instability.

The paper is organized as follows.
 In Sec. II, we characterize the stationary helical shear flow from its vorticity distribution in the Gross-Pitaevskii (GP) model at zero temperature.
 Sections III and IV discuss the linear stability of the helical shear flows and show an instability phase diagram on the basis of an effective theory and numerical analyses of the Bogoliubov theory.
 In Sec. IV, we demonstrate the nonlinear development of the helical shear-flow instability by numerically solving the GP equation.
 Section V is devoted to the conclusion and discussion.

\section{Stationary helical shear flow}
Let us consider two-component BECs in a uniform system in the GP model at zero temperature \cite{PethickSmith}.
 Two-component BECs are well described by the complex order parameter ${\Psi_j}({\bm r},t)$ of the $j$th component.
 The mean-field Lagrangian is given by
\begin{eqnarray}
{\cal L}=\int d^3x\[ {{\cal{P}}_{1}}+{{\cal{P}}_{2}}-g_{12}|\Psi_1|^2|\Psi_2|^2\]
\label{lag_1}
,\end{eqnarray}
 with
\begin{eqnarray}
{\cal{P}}_{j}=i\hbar{\Psi^*_j}{\partial_t}{\Psi_j}-{\frac{\hbar^2}{2{m_j}}}|{\bm \nabla}{\Psi_j}|^2-\frac{g_{jj}}{2}|\Psi_j|^4+\mu_j|\Psi_j|^2,
\label{pressure}
\end{eqnarray}
 where $m_j$ and $\mu_j$ are the particle mass and the chemical potential of the $j$th component, respectively.
 The interaction parameters $g_{jk}$ are defined as ${g_{jk}}=2{\pi}{\hbar^2}{a_{jk}}({m_j}+{m_k})/{m_j}{m_k}$ $(j,k=1,2)$
 with the $s$-wave scattering length $a_{jk}$ between the $j$th and $k$th components.
 The parameters are assumed to satisfy the condition for phase separation, ${g_{12}}/\sqrt{{g_{11}}{g_{22}}}>1$.
 In the following, we shall consider experimentally feasible situations $m=m_1=m_2$, $g=g_{11}=g_{22}$, and $g_{12}/g=1.2,$ e.g., weakly segregated condensates in \Ref{TojoPRA2010}.
Considering stationary states with rotational symmetry along the $z$ axis,
 ${\Psi_j}({\bm x},t)=\Phi_j({\bm x})\equiv{\sqrt{n_j(r)}}e^{i\({Q_j}z+L_j\phi \)}$
 with the cylindrical coordinates ${\bm x}=(r, \phi, z)$,
 one obtains the time-independent GP equations
\begin{eqnarray}
\(h_j+\sum_k g_{jk}n_k-\mu_j\)\Phi_j=0.
\label{GPeqs}
\end{eqnarray}
 Here, we used $h_j=-\frac{\hbar^2}{2m_j}\(\p_r^2+r^{-1}\p_r-Q_j^2-L_j^2/r^2\)$.
 Because we are interested in the vortex state illustrated in \Fig{fig:schematic}, we set without loss of generality
\bea
(Q_1,~Q_2,~L_1,~L_2)=(0,~Q,~L,~0),
\label{QuantumNumber}
\eea
where the first component has a linear velocity $\frac{\hbar Q}{m}$ parallel to the $z$ axis and the second component has a rotational velocity $\frac{\hbar L}{m r}$ around the axis.
In the following, we use the dimensionless parameters $\gamma=g_{12}/g$ and $\nu\equiv \mu_2'/\mu_1$ with $\mu_2'=\mu_2-\frac{\hbar^2Q^2}{2m_2^2}$
 and the scales $\xi=\sqrt{{\hbar^2}/(m{\mu_1})}$, $\tau=\hbar/{\mu_1}$, and $c=\sqrt{{\mu_1}/m}$ for length, time, and velocity, respectively.

\begin{figure}[h!]
 \centering
 \includegraphics[width=.9 \linewidth]{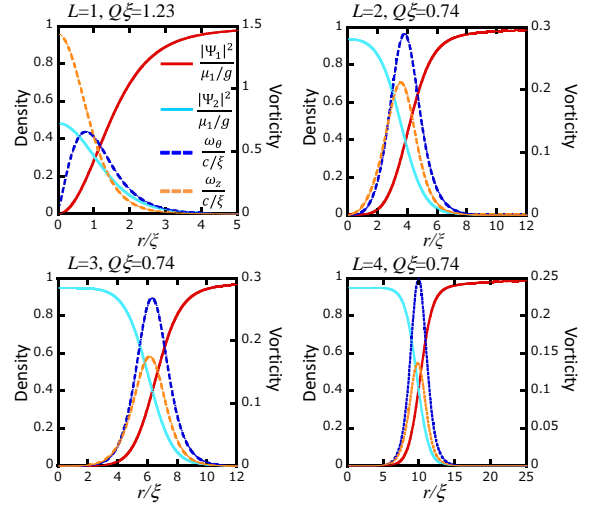}
 \caption{
 (Color online) Radial profiles of the density of two components $|\Phi_j|^2$ and vorticity ${\bm w}$ in the stationary state
 with $\gamma=0.95$ and $\nu=1.2$.
 The radius $r_0$ of the interface, defined by the plane $|\Phi_1|=|\Phi_2|$, increases with $|L|$.
 The horizontal axis is the radial coordinate.
 The vertical axes on the left and right hand sides show the normalized density $\left(\frac{|\Psi_1|}{\mu_1/g},~ \frac{|\Psi_1|}{\mu_1/g}\right)$ and the vorticity $\left(\frac{\omega_\theta}{c/\xi},~\frac{\omega_z}{c/\xi}\right)$, respectively.
}
 \label{fig:dens_vor}
\end{figure}

Figure \ref{fig:dens_vor} shows the radial profiles of characteristic helical shear flows in the stationary states with $\nu=0.95$.
 When the first component contains multiquantized vortices [$L=4$ in \Fig{fig:dens_vor}],
 the centrifugal force resulting from the rotational flow in the first component enlarges the radius $r_0$ of the circular interface plane,
 defined by the plane $|\Psi_1|=|\Psi_2|$.
The interface radius $r_0$ becomes smaller as $L$ decreases.
The interface is ill defined for $L=1$ in \Fig{fig:dens_vor}(d),
where the radius $r_0$ becomes comparable to the interface thickness $\sim\xi$.

To classify the stationary shear-flow states and characterize their hydrodynamic aspects,
 it is convenient to introduce the mass-current velocity ${\bm v}$ and the vorticity ${\bm w}$, defined as
\begin{eqnarray}
{\bm v}&=&\frac{\frac{\hbar}{2i}\sum_j \[ \Psi_j^*({\bm \nabla}\Psi_j)-({\bm \nabla}\Psi_j^*)\Psi_j\]}{\sum_j m_j |\Psi_j|^2},
\label{mass_vel}\\
{\bm w}&=&{\bm \nabla}\times {\bm v}.
\end{eqnarray}
 For $L=4$ in \Fig{fig:dens_vor}, the velocity ${\bm v}$ changes sharply and the vorticity ${\bm w}=(0,~w_\phi,~w_z)$ is localized around $r=r_0$,
 where the vortex sheet forms a circular pipe from the hydrodynamic viewpoint.
 Since the trajectory along the direction of ${\bm w}$ is helical, this state may be called a helical vortex sheet.
 When the interface radius $r_0$ becomes close to the $z$ axis ($r=0$) for $L=1$ in \Fig{fig:dens_vor},
 the density $n_2(r=0)$ at the center is suppressed so that the vorticity is distributed around $r=0$,
 forming a linear vortex with a superflow along the vortex core.
 We call such a vortex state a core-flow vortex.

 The interface radius $r_0$ depends on the particle populations of the two components.
 Because the populations of the first and second components are  decreasing and increasing functions of $\nu$ with $L$ fixed, respectively,
 the radius $r_0$ is an increasing function of $\nu$.
 There exists a critical ratio ($\nu_{\min}$) of $\nu$,
 below which the population of the second component vanishes in the vortex core.
 In the limit of $\Phi_2 \to 0$ for $\nu \to \nu_{\min}$,
 by neglecting the   $n_2$ terms in \Eq{GPeqs},
 $\Phi_2$ corresponds to the single-particle wave function of the ground state in an external potential $g_{12}n_1$.
 Consequently,  the value $\nu_{\min} \mu_1$ equals the ground-state energy;
 e.g., for $L=1$, we have $\nu_{\min}=a_1\sqrt{\gamma}\sim 0.899$ with $n_1=\frac{\mu_1}{g}a_1^2(r/\xi)^2$ ($r\to 0)$.
 However, we found that the interface radius $r_0$ is larger than the system size for $\nu \to 1$ in the numerical simulations.
 This fact may be understood from the hydrodynamic viewpoint discussed in the  next section: the small difference between the hydrostatic pressures on the two sides of the interface balance for sufficiently large $r_0$.
 Although the instability discussed below also depends on the ratio $\nu$,
 we consider the instability for $\nu=0.95$ in the following;
  this case is sufficient to show the characteristic behavior of the shear-flow instability.

\section{Effective theory of helical KHI}
First, we perform a linear stability analysis of the helical vortex sheet.
 The effective theory of quantum KHI of a flat vortex sheet \cite{2010TakeuchiPRB, phys_re} is applied to our case of a helical vortex sheet.
 In the effective theory, we consider only the radial shift $\eta$ of the interface position and neglect the thickness.
 The position of the interface is represented by  a curve $r={r_0}+\eta({\phi},z,t)$.

In this approximation, the Lagrangian [\Eq{lag_1}] is reduced with the interface-tension coefficient $\alpha$ and the interface area $S$ to
\begin{eqnarray}
&&{\cal L}_{\rm eff}=\int^\infty_{-\infty} dz \int ^\pi_{-\pi} d\phi
\(\int^{\infty}_{r_0+\eta} rdr {\cal P}_1+\int^{r_0+\eta}_{0} rdr {\cal P}_2\)
\nonumber\\
&& \ \ \ \ \ \ \ \ \ \ \ \ -{\alpha}S.
\label{lagrangian2}
\end{eqnarray}
 The area $S$ equals $S_0=2\pi \int dz r_0$ in the stationary sate with $\eta=0$.
 For a small deformation, we have
\begin{equation}
S{\approx}S_0 +\int dz d\phi \[\eta+ \frac{1}{2r_0}\(\p_\phi \eta\)^2+\frac{r_0}{2}\(\p_z \eta\)^2 \].
\label{surface area}
\end{equation}
 The variation of \Eq{lagrangian2} with respect to $\eta$ gives
\begin{eqnarray}
{\lefteqn{-({r_0}+\eta){{\cal{P}}_{1}({r_0}+\eta)}+({r_0}+\eta){{\cal{P}}_{2}({r_0}+\eta)}}{\nonumber}} \\
&+&{\alpha}\left\{-1+{r_0}\left({\frac{1}{{r_0}^2}}{{{\partial_\phi}^2}\eta}+{{{\partial_z}^2}\eta}\right)\right\}=0.
\label{Bernoulli}
\end{eqnarray}
 This equation shows the pressure equilibrium on the interface in the stationary state
\begin{equation}
P_2-P_1=\frac{\alpha}{r_0},
\label{equilibrium condition}
\end{equation}
 where $P_j$ is the value of \Eq{pressure} along the interface in the stationary state, called the hydrostatic pressure.
 The left- and right-hand sides of \Eq{equilibrium condition} refer to the pressure difference across the interface
 and the tension resulting from the curvature of the interface, respectively.
 In the Thomas-Fermi approximation neglecting the quantum pressure term $\propto (\bm{\nabla} n_j)^2/n_j$ in \Eq{pressure},
 one obtains $P_j=\frac{1}{2}gn_j^2(r_0)$ with $n_1= \[\mu_1-\frac{\hbar^2L^2}{2mr^2} \]/g$ and $n_2= \mu'_2/g$.

\begin{figure}[htbp]
 \centering
 \includegraphics[width=.9 \linewidth]{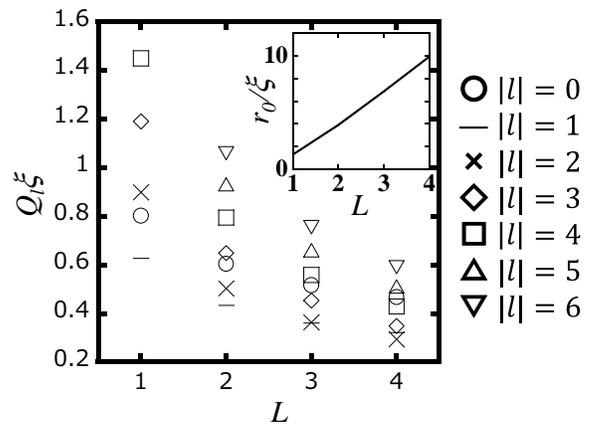}
 \caption
{
 $L$ dependence of $Q_l$ in \Eq{cvelocity} for $1 \leq L \leq 4$ with $\nu=0.95$ and $\gamma=1.2$.
The insert shows the $L$ dependence of the interface radius $r_0$ in \Eq{radius}.
}
 \label{V_l}
\end{figure}

The equilibrium radius $r_0$ is calculated from \Eq{equilibrium condition}.
 For large $r_0$, one can take the coefficient $\alpha$ in \Eq{equilibrium condition} to be that for a flat interface, calculated by assuming $P_1\to P_2$ \cite{Ao1998,Schaeybroeck2008},
\begin{equation}
{\alpha}=\alpha_0\nu^{\frac{3}{2}}
,\end{equation}
 with the tension coefficient $\alpha_0=\frac{\xi \mu_1^2}{\sqrt{2}g} \sqrt{\gamma-1}$ under the external pressure $\mu_1^2/2g$.
 Neglecting the term of $o\left(L^4\xi^4/r_0^4\right)$,
 we obtain
\begin{equation}
\frac{r_0}{\xi}=\frac{\sqrt{\frac{\nu^3(\gamma-1)}{2}+L^2\(1-\nu^2\)}-\sqrt{\frac{\nu^3(\gamma-1)}{2}}}{1-\nu^2}.
\label{radius}
\end{equation}
 This equation holds for $1/\sqrt{2}< r_0/L\xi <1/\sqrt{2(1-\nu)}$ from the conditions $n_1(r_0)>0$ and $P_2>P_1$.
 The expression \eq{radius} is in good agreement with the numerical results of the interface radius of the helical vortex sheets in \Fig{fig:dens_vor} [see also the insert in \Fig{V_l}],
 and thus, we can compare the stability analysis demonstrated below with the numerical analysis presented in the next section.

Linear stability is investigated by considering a small fluctuation
\begin{eqnarray}
|\Psi_j({\bm x},t)|^2&=&n_j(r)+\delta n_j({\bm x},t), \label{linear1} \\
\arg\Psi_j({\bm x},t)&=&Q_jz+L_j\phi+\delta\theta_j({\bm x},t), \label{linear2}
\end{eqnarray}
 with
\begin{eqnarray}
{\eta}&=&{\bar{\eta}}{\ }{\sin}(l{\phi}+qz-{\omega_{l,q}}t) \label{delta eta}, \\
{\delta}{n_{j}}&=&{\delta}{{\bar n}_j}(r){\sin}(l{\phi}+qz-{\omega_{l,q}}t) \label{delta n}, \\
{\delta}{\theta_{j}}&=&{\delta}{{\bar{\theta}}_{j}}(r){\cos}(l{\phi}+qz-{\omega_{l,q}}t), \label{delta phi}
\end{eqnarray}
 where $l$ and $q$ are real.
 The fluctuation represents the ripple mode, which causes a ripple wave on the interface.
 The kinematic boundary condition is employed along the interface
\bea
\[{\bm v}_j\]_r=\partial_t \eta+{\bm v}_j\cdot{\bm \nabla}\eta,
\label{bound}
\eea
 with the radial component $\[{\bm v}_j\]_r$ of the local superfluid velocity
 ${\bm v}_j=\frac{\hbar}{2im|\Psi_j|^2}\[ \Psi_j^*({\bm \nabla}\Psi_j)-({\bm \nabla}\Psi_j^*)\Psi_j\]$ of the $j$th component.
 The radial shift of the interface causes a density fluctuation along the interface,
 $\delta n_1=-\eta\partial_r n_1=-\eta\frac{\hbar^2L^2}{mgr_0^3}$ and $\delta n_2=-\eta\partial_r n_2=0$.
 The boundary condition \eq{bound}, together with the equation of continuity,
 $\partial_t \delta n_j+{\bm \nabla}(n_j\delta{\bm v}_j)+{\bm \nabla}(\delta n_j {\bm v}_j)=0$,
 yields $\[{\bm \nabla}\cdot \delta{\bm v}_j\]_{r=r_0}=0$ with $\delta{\bm v}_j=\frac{\hbar}{m_j}{\bm \nabla}\delta\theta_j$.
 Since the fluctuation is considered to be localized around the interface,
 we choose a form of the phase fluctuation as
\bea
\delta{\bar{\theta}}_1(r)&=&{c_1}{K_{l}(|q|r)}, \label{bessel1} \\
\delta{\bar{\theta}}_2(r)&=&{c_2}{I_{l}(|q|r)}, \label{bessel2}
\eea
 where $c_1$ and $c_2$ are constants and $I_l$ and $K_l$ are modified Bessel functions of the second and first kinds, respectively.

By linearizing \Eq{Bernoulli} together with Eqs. \eq{bound}, \eq{bessel1}, and \eq{bessel2},
 we obtain the dispersion relation
\bea
\omega_{l,q}=\frac{\Xi_1\frac{\hbar lL}{mr_0^2}+\Xi_2\frac{\hbar qQ}{m}+\sqrt{A}}{\Xi_1+\Xi_2}
\label{dispersion}
,\eea
 with
\bea
A&=&\(\Xi_1+\Xi_2\)|q|\[f+\alpha\(\frac{l^2}{r_0^2}+q^2\)\]
\nonumber\\
&&-\Xi_1\Xi_2\({\bm q}\cdot{\bm V}_R\)^2.
\label{root}
\eea
 Here, we used
 $\Xi_1=\frac{2mn_1(r_0)K_l(|q|r_0)}{K_{l-1}(|q|r_0)+K_{l+1}(|q|r_0)}$,
 $\Xi_2=\frac{2mn_2(r_0)I_l(|q|r_0)}{I_{l-1}(|q|r_0)+I_{l+1}(|q|r_0)}$,
 $f=\frac{{\hbar^2}{L^2}}{{m}{r_0^3}}{n_1(r_0)}-\frac{\alpha}{r_0^2}$,
 ${\bm q}=(0,l/r_0, q)$,
 and ${\bm V}_R=\frac{\hbar}{m}(0,L/r_0,-Q)$.
 The frequency $\omega_{l,q}$ is real for $A\geq 0$ but becomes complex for $A<0$.
 For $A<0$, the helical vortex sheet is dynamically unstable against some perturbation
 so that the ripple modes are amplified exponentially with time.
 For $A\geq 0$, the $\hbar\omega_{l,q}$ may correspond to the energy quanta of ripple modes or {\it ripplons}.
 In the presence of energy dissipation,
 the helical vortex sheet is thermodynamically unstable when a ripple mode has negative energy with $\omega_{l,q}<0$.
 Here, we restrict the discussion to dynamic instability by neglecting thermodynamic instability,
 which is experimentally reasonable in atomic BEC systems.

The dispersion relation \eq{dispersion} has the generalized form of that of quantum KHI \cite{2010TakeuchiPRB}
 and capillary instability \cite{Sasaki} in phase-separated two-component BECs.
 For $L=Q=l=0$, where there is no shear flow between the two components,
 the dispersion relation \eq{dispersion} reduces to those of the Plateau-Rayleigh (or capillary) instability \cite{Christiansen}.
 In general, a capillary flow, namely, a helical vortex sheet with $L=0$, is dynamically unstable because $A<0$ with $f=-\alpha/r_0^2$ in \Eq{root},
whereas a helical vortex sheet with $L\neq 0$ can be stabilized because of the centrifugal force in the presence of a vortex at $r=0$.
 However, in the limit of $r_0\to \infty$,
 the dispersion relation \eq{dispersion} reduces to that of  the quantum KHI for a flat interface without external potentials in \Ref{2010TakeuchiPRB}.
 In this sense, the instability for the case with a sufficiently large $r_0$ may be called the helical KHI.
 In the same manner as in those instabilities,
 our effective theory is based on the assumption that the fluctuations are well localized around the interface
 and the wavelength of the ripple modes are smaller than the thickness of the interface.

The phase diagram for the linear stability is calculated from the dispersion relation \eq{dispersion}.
 The frequency \eq{dispersion} of a ripple mode with $l$ becomes complex when $Q$ exceeds the critical value
\begin{eqnarray}
Q_l=\min_q (Q_{l,q})
\label{cvelocity}
,\end{eqnarray}
with
\bea
Q_{l,q}=\frac{1}{|q|}\left| \frac{m}{\hbar}\sqrt{|q|\frac{\Xi_1+\Xi_2}{\Xi_1\Xi_2}\[f+\alpha\(l^2/r_0^2+q^2\)\]} -\frac{L|l|}{r_0^2}\right|.
\nonumber
\eea
 We plotted the $L$ dependence of $Q_l$ with $\nu=0.95$ and $\gamma=1.2$ for $0 \leq l \leq 6$ in \Fig{V_l}.
 The helical shear flow becomes dynamically unstable when the velocity $\frac{\hbar}{m}Q$ of the second component exceeds the critical value
\bea
V_c=\frac{\hbar}{m}\min_l(Q_l).
\label{V_c}
\eea
 The critical velocity $V_c$ decreases monotonically with $L$ as seen in \Fig{V_l}.
 The velocity $V_c$ approaches zero asymptotically   for $L\to \infty$
 since then the interface radius $r_0$ increases with $L$ as $\sim\xi L/\sqrt{1-\nu}$ and then the stabilizing force $f$ is asymptotic to zero.

\section{Crossover from helical KHI to  core-flow instability }
When the interface radius $r_0$ becomes small and the helical vortex sheet turns into a core-flow vortex,
 the ripple modes are ill defined, and thus, the effective theory  can no longer be applied.
 To determine the stability in the crossover regime between the helical vortex sheets ($r_0\to \infty$) and the core-flow vortices ($r_0\to 0$),
 we demonstrate numerically the regime's linear stability based on  Bogoliubov theory \cite{PethickSmith}.

 By introducing a perturbation of the order parameters
\bea
\delta\Psi_j=e^{i(L_j\phi+Q_jz)}\( {\cal U}_j-{\cal V}_j^* \)
,\eea
 and linearizing the time-dependent GP equations
\bea
i\hbar \partial_t \Psi_j=\(-\frac{\hbar^2}{2m_j}{\bm \nabla}^2+\sum_k g_{jk}|\Psi_j|^2-\mu_j\)\Psi_j,
\label{timeGP}
\eea
 around the stationary states $\Psi_j=\Phi_j$ with ${\cal U}_j=u_j(r)e^{i(qz+l\phi-\omega t)}$ and ${\cal V}_j=v_j(r)e^{i(qz+l\phi-\omega t)}$,
 we obtain the Bogoliubov--de Gennes (BdG) equations
\begin{equation}
{\hbar}{\omega}{\bm u}=\sigma_3{\cal{H}}{\bm u},
\label{BdG}
\end{equation}
 where we used ${\bm u}=(u_1,u_2,v_1,v_2)^T$, $\sigma_3={\rm diag}(1,1,-1,-1)$,
\begin{eqnarray}
{\cal{H}}=
\begin{pmatrix}
h_{1}^{+}+{\cal G}_{11}                & {\cal G}_{12} & {\cal G}_{11}             & {\cal G}_{12} \\
{\cal G}_{12} & h_{2}^{+}+{\cal G}_{22}                & {\cal G}_{12} & {\cal G}_{22} \\
{\cal G}_{11}              & {\cal G}_{12} & h_{1}^{-}+{\cal G}_{11}                & {\cal G}_{12} \\
{\cal G}_{12} & {\cal G}_{22}              & {\cal G}_{12} & h_{2}^{-}+{\cal G}_{22}
\end{pmatrix}
\label{matrix}
\end{eqnarray}
 with ${\cal G}_{jk}=g_{jk}\sqrt{n_jn_k}$, and
\bea
h_j^{\pm}&=&-\frac{\hbar^2}{2m_j}\[ \p_r^2+\frac{1}{r} \p_r-\frac{(l\pm L_j)^2}{r^2}-(q\pm Q_j)^2\]
\nonumber \\
&&  -\mu_j+\sum_k g_{jk}n_k.
\label{h_jpm}
\eea
 An eigenvalue $\omega$ of \Eq{BdG} represents the frequency of an elementary excitation in the helical shear-flow state.
 Because a solution $(\omega,q, l, u_j, v_j)$ has its conjugate solution $(-\omega^*,-q, -l, v_j^*, u_j^*)$,
 we present here only the results for $q\geq 0$ without loss of generality.
\begin{figure}[htbp]
 \centering
 \includegraphics[width=1.0 \linewidth]{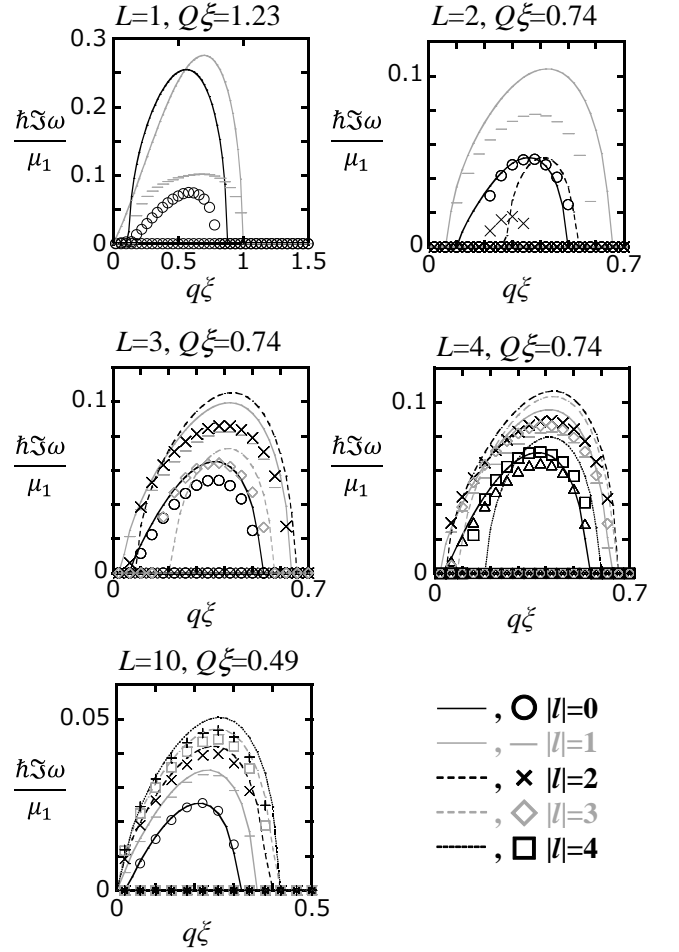}
 \caption
{
 The imaginary part ${\rm Im} \omega  \geq 0$ for several values of $l$ obtained by the analytical (curves) and numerical (points) calculations based on the effective theory and the Bogoliubov theory, respectively.
 We set the parameter $Q\xi=1.23$ for $L=1$ and $Q\xi=0.74$ for $L=2, 3, 4$ with $\nu=0.95$ and $\gamma=1.2$.
}
 \label{fig:diss_com}
\end{figure}
\begin{figure}[htbp]
 \centering
 \includegraphics[width=1.0 \linewidth]{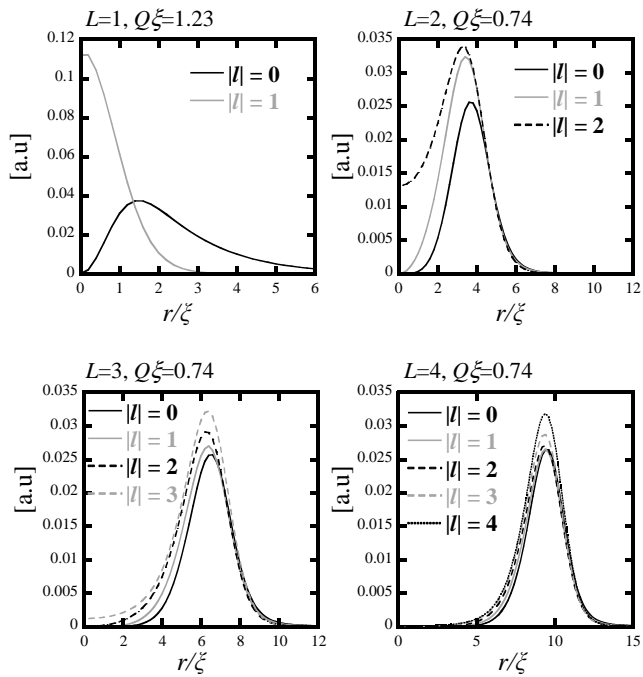} 
 \caption
{
 The radial distribution of $|\delta\Psi_1(r,\phi, z, t)|^2$ of the first component for $\phi=z=t=0$ obtained by numerically solving the BdG equations \eq{BdG}.
  The splitting modes ($|l|=L$) have a finite value at $r=0$ while the modes with $|l|\neq L$ are distributed locally around the interface $r=r_0$.
 We set the parameter $Q\xi=1.23$ for $L=1$ and $Q\xi=0.74$ for $L=2, 3, 4$ with $\nu=0.95$ and $\gamma=1.2$.
}
 \label{fig:ExD}
\end{figure}

Figure \ref{fig:diss_com} shows the distribution of the imaginary part ${\rm Im} \omega >0$ for several values of $l$ obtained
 by numerically solving the BdG equations \eq{BdG}.
 The effective theory well explains the imaginary part of the the numerical results for all $l$ in the limit $r_0\to \infty$ (see, e.g., the plot for $L=10$ in \Fig{fig:diss_com}).
 However, in the crossover regime where the interface radius $r_0$ becomes comparable to the interface thickness $\sim \xi$,
 the imaginary part for the specific mode $|l|=L>1$ of the numerical results deviates from
 that of the analytic result compared to the other modes, as is seen in \Fig{fig:diss_com} for $L>1$.
 The deviation is related to a specific behavior in the spatial distribution of the $|l|=L$ mode.
 We can see that the distribution $|\delta\Psi_1|^2$ has a finite value at $r=0$ only for $|l|=L$ (see \Fig{fig:ExD}).
 This is because
 the amplitude of $u_1$ ($v_1$) can survive at $r=0$ for $l=-L$ ($l=L$) with the third term in \Eq{h_jpm} being zero.
 However,
 the modes with $|l|\neq L$ are distributed locally around the interface $r=r_0$ in \Fig{fig:ExD},
 and then the effective theory for the ripple modes is relatively valid for describing the imaginary part ${\rm Im} \omega(q)$ in \Fig{fig:diss_com}.

The instability reflects the properties as a quantized vortex rather than those as a vortex sheet
 when the interface radius $r_0$ becomes small enough.
 In general, since the intercomponent interaction is essential to the shear-flow instability,
 it is expected that the instability is suppressed when the population of the second component is small (see $L=1$ in \Fig{fig:dens_vor}).
 Actually, the imaginary part of the numerical results is substantially smaller than that of the analytic estimation for $L=1$ in \Fig{fig:diss_com}.
 The instability vanishes for $L=1$ in the limit $r_0\to 0$ for $\nu \to \nu_{\min}$ without the second component.
 Note that, for $|L|> 1$, a dynamic instability can survive even in the limit $r_0\to 0$.
 The survived instability is well known as the vortex-splitting instability,
 which causes the splitting of a multiquantized vortex into single-quantized vortices \cite{2003Mottonen, 2004Kawaguchi}.
 The abnormal behavior of the $|l|=L$ mode above is regarded to be a reflection of the splitting instability.
 In this sense, we call the $|l|=L$ mode the splitting mode in this paper.

The instability of the core-flow vortex with $L=1$ is triggered by the vortex-characteristic modes: Kelvin ($|l|=1$) and varicose ($l=0$) modes.
 If the velocity of the second component or $Q$  becomes smaller,
 only the Kelvin mode ($|l|=1$) has a nonzero imaginary part: ${\rm Im} \omega\neq 0$.
 The effect of the Kelvin mode is qualitatively different from that of the ripple modes in the helical vortex sheet with larger $L$;
 although the ripple modes localized around $r=r_0$ do nothing other than cause ripples to propagate along the interface,
 the Kelvin mode makes the position of the core slide from $r=0$.
 When $Q$ is increased,
 the varicose mode ($l=0$) begins to have complex frequencies in addition to the Kelvin mode in the core-flow vortex with $L=1$.
 The varicose mode induces density perturbations with rotational symmetry around the vortex axis in the density distribution $|\Psi_j|^2$.
 The perturbations may be considered as oscillations of the radius of the vortex core since the varicose mode causes the $|\Psi_1|=|\Psi_2|$ plane to oscillate.
 We call this dynamic instability, triggered by the Kelvin and varicose modes in the core-flow vortex with $L=1$,
  the core-flow instability (CFI) for the purpose of distinguishing it from the helical KHI in the helical vortex sheet regime.

\begin{figure}[htbp]
 \centering
 \includegraphics[width=.9 \linewidth]{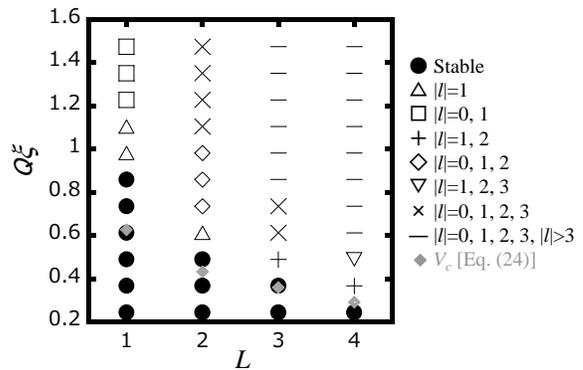}
 \caption
{
 Phase diagram of the helical shear-flow instability for $\gamma=1.2$ and $\nu=0.95$.
 The stable region is shown by filled circles .
 Different marks represents which modes can be amplified in the instability development.
 Filled diamonds are the analytical results in \Eq{V_c}.
}
 \label{fig:phase_diagram4}
\end{figure}

We summarize here the linear stability of the helical shear-flow states with a phase diagram in \Fig{fig:phase_diagram4}.
 The helical shear-flow states are dynamically stable below the critical value of $Q$ or the critical velocity of the second component.
 It is surprising that the analytic result of the critical velocity ($V_c$) in \Fig{V_l} is in good agreement with the numerical result
 even in the crossover regime with small $L>1$ in \Fig{fig:phase_diagram4}.
 This accidental agreement comes from the fact that the critical value $(Q_{l=|L|})$ of the splitting mode is higher than that of the ripple modes,
 which is described well with the effective theory.
 Different modes have complex frequency for larger value of $Q,$ leading to a more complicated instability.
 The simplest case is the core-flow vortex ($L=1$),
 where the instability is caused mainly by Kelvin or varicose modes.

\section{Nonlinear development}
To gain deeper insights into the helical shear-flow instability,
 let us discuss the nonlinear development of the instability phenomena by numerically solving the GP equations \eq{timeGP}.
 We restrict our investigation to the cases of the crossover or core-flow vortex regimes
 since the nonlinear development in the helical KHI limit ($r\to r_0$) must be essentially the same as that of the quantum KHI
 for a flat vortex sheet discussed in \Refs{2010TakeuchiPRB,2010SuzukiPRA}.

The numerical simulations were performed on a periodic system along the $z$ axis.
 Similar situations can be realized experimentally by considering, e.g., elongated condensates in a cigar-shaped potential.
 To neglect the finite-size effect and capture the essence of the instability,
 we investigate the developments without a potential gradient around the $z$ axis
 by using the cylinder potential $V_j=V_{\rm trap}\tanh [ (r^2-R_{\rm trap}^2)/\xi^2]$
 with $V_{\rm trap}= 50\mu_1$ and $R_{\rm trap}\gg \xi$.
 Unless otherwise noted,
 all the time development diagrams displayed in the following are obtained by using numerical simulations starting
 from  stationary states $\Phi_j$ with a small amount of white noise added  to trigger the dynamic instability,
 which breaks the translational and rotational symmetry along the $z$ axis.
 The time evolution is obtained by solving the GP equation with the Crank-Nicolson method.
 The parameters of the initial states are fixed as $\gamma=1.2$ and $\nu=0.95$.

As a first step, we demonstrate the instability development from the core-flow vortex ($L=1$).
 This instability corresponds to   a CFI triggered by the Kelvin and varicose modes,
 which is distinguishable from the instabilities by the amplifications of the splitting and the ripple modes.
 We discuss the instability for the case of $L>1$ after the CFI.

\subsection{CFI by Kelvin mode amplification ($L=1$)}

\begin{figure}[htbp]
 \centering
 \includegraphics[width=1.0 \linewidth]{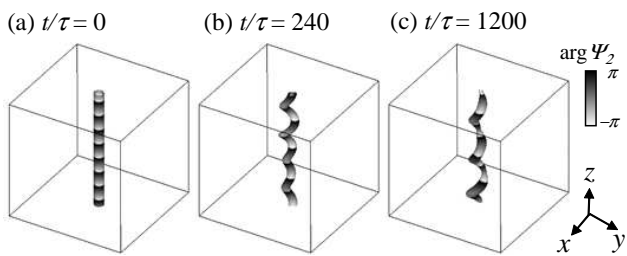}
 \caption
{
 Nonlinear development of the Kelvin mode amplification in the CFI of the core-flow vortex for $L=1$ and $Q\xi=0.98$.
 The surface shows the isosurface $|{\Psi}_1|^2=|{\Psi}_2|^2$ and the surface shading represents the phase of the second component on the isosurface.
 The system size is $25.6\xi$ in the $x$ and $y$ directions and the periodicity is $51.2\xi$ in the $z$ direction.
}
 \label{CFIbyKelvin}
\end{figure}

The simplest case is  the CFI developing from a core-flow vortex,
 where only the Kelvin modes have complex frequency.
 Figure \ref{CFIbyKelvin} shows the typical development of the  CFI owing to  Kelvin mode amplification for $Q\xi =0.98$.
 The dynamic instability amplifies mainly the Kelvin mode with the largest value of the imaginary part ${\rm Im} \omega$ in this system,
 which is obtained by  numerically  solving the BdG equation.
 The amplification leads to a deformation of the initial straight vortex line [\Fig{CFIbyKelvin}(a)] into a helix [\Fig{CFIbyKelvin}(b)] with a periodicity that corresponds to the wavelength of the most amplified Kelvin mode.
 The radius of the helix grows with time and the growth stops up to a finite radius.
 After the growth, the core-flow vortex  typically keeps its helical structure
 although the helix is somewhat deformed [\Fig{CFIbyKelvin}(c)].

The direction of the helix appearing in the CFI is uniquely determined.
 To understand this fact qualitatively, let us parametrize the trajectory ${\bm s}$ of the helix with $z$ as
\bea
{\bm s}(z)=[r_s \cos(k_s z),r_s\sin(k_s z),z]
\eea
 in Cartesian coordinates.
 Here, $r_s$ and $k_s$ are the radius and the wave number of the helix, respectively.
 It is important to note that the Kelvin mode amplification reduces the angular momentum of the first component containing a vortex
 because the amplification leads to a displacement of the vortex core from the $z$ axis.
 During this process, the angular momentum of the second component must increase
 since the sum of the angular momenta of the two components is conserved in our system with  rotational symmetry about the $z$ axis.
 In the initial straight vortex state of \Eq{QuantumNumber} with $Q>0$ and $L>0$,
 the first component has a positive angular momentum about the $z$ axis.
 Since then the angular momentum of the second component becomes positive for $k_s>0$ and negative for $k_s<0$,
 the helix must be right-handed with $k_s>0$ to conserve  angular momentum.
 Therefore, whether the helix becomes right-handed ($k_s>0$) or left-handed ($k_s<0$) is determined by the signs of $Q$ and $L$;
 e.g., $k_s<0$ for the core-flow vortex with $Q<0$ and $L>0$.

Since there remains a relative velocity along the vortex line after the helical deformation,
 one may expect the CFI to occur additionally at the local point along the core-flow vortex.
But this is not the case.
 Note that, in \Fig{CFIbyKelvin}, the winding number of the phase $\arg\Psi_2$ of the second component displayed is kept through the instability dynamics.
 In that case, the current velocity of the second component parallel to the helical trajectory is estimated
 as $\frac{\hbar}{m_2}Q/\sqrt{1+k_s^2r_s^2}$.
 Thus, the relative velocity along the vortex core decreases with the radius $r_s$ of the helix.
 Consequently, the helical deformation works to suppress the additional instability caused by the relative velocity along the vortex core.

The growth dynamics of a helical vortex line in the CFI is similar to that in the Kelvin wave instability \cite{2009Takeuchi}.
 However, the CFI is essentially a different phenomenon from the Kelvin wave instability.
 The CFI belongs to a dynamic instability caused by an {\it internal} interaction,
 the intercomponent interaction between the two components,
 and thus, the sums of the momenta and the angular momenta along the $z$ axis of the two
 are, respectively, conserved in our system with  translational and rotational symmetry about the axis.
 In contrast, those quantities are not conserved in the Kelvin wave instability,
 where the thermodynamic instability is caused by  energy dissipation resulting from {\it external} interaction with the environment.
 This essential difference between the two instabilities appears in their typical nonlinear developments;
 the radius of the helix stops  growing within a finite range in the CFI,
 but, in the Kelvin wave instability, the radius continues to increase monotonically unless the vortex collides with other vortices or the system reaches  a local minimum of energy.

\subsection{Influence of varicose motion ($L=1$)}
\begin{figure}[htbp]
 \centering
 \includegraphics[width=1 \linewidth]{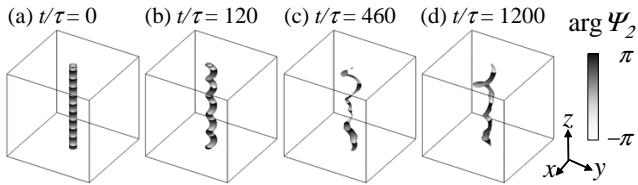}
 \caption
{
 Nonlinear development of the CFI induced by Kelvin and varicose modes for $L=1$ and $Q\xi=1.23$.
 The system size is $25.6\xi$ in the $x$ and $y$ directions and the periodicity is $51.2\xi$ in the $z$ direction.
}
 \label{kelvinVSvaricose}
\end{figure}

When $Q$ is increased, varicose modes begin to have complex frequencies.
 Figure \ref{kelvinVSvaricose} shows the nonlinear development of the CFI for $Q\xi=1.23$,
 where both Kelvin and varicose modes have complex frequencies.
 In addition to the helical deformation of the vortex core resulting from  Kelvin mode amplification [\Fig{kelvinVSvaricose}(b)],
  varicose mode amplification causes oscillations of the radius of the vortex core [\Fig{kelvinVSvaricose}(c)].
 This oscillation comes mainly from density modulation of the second component trapped along the vortex core.
 We found that the winding number of the phase $\arg\Psi_2$ across the system in the $z$ direction changes
 from \Fig{kelvinVSvaricose}(a) to \Fig{kelvinVSvaricose}(d).
 After the relative velocity decreases because of the reduction of the winding number,
 oscillations of the core radius are essentially suppressed but the helical deformation of the vortex core survives.

\begin{figure}[htbp]
 \centering
 \includegraphics[width=1.0 \linewidth]{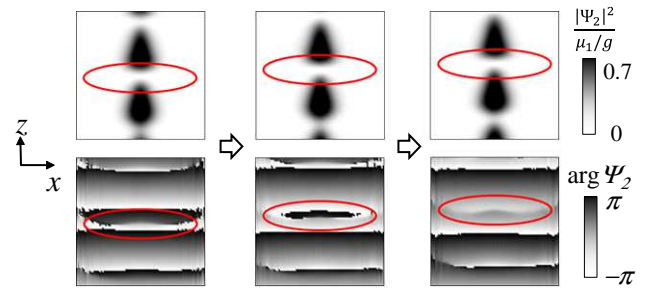}
 \caption
{
 (Color online) Time development of the second component in the CFI from varicose modes for $L=1$ and $Q\xi=1.47$.
 Top and bottom panels show the close-up longitudinal profiles ($y=0$) of the density $|\Psi_2|^2$ and the phase $\Theta_2=\arg\Psi_2$ along the vortex core.
 (a) The varicose mode amplification leads to  density troughs in the density profile along a narrow channel made by the vortex of the first component.
 The phase gradient is steep in the density troughs.
 (b) Two vortices from the outside approach each other around the trough as seen in the phase $\Theta_2$.
 (c) After the vortices annihilate, the phase becomes smooth and the relative velocity between the two components is decreased.
}
 \label{varicose}
\end{figure}

To better understand how the varicose mode amplification causes the reduction of the winding number in the second component,
 we examined the instability development starting from the stationary state with an initial perturbation, which does not break rotational symmetry around the $z$ axis.
 This perturbation results purely in the amplification of varicose modes without  Kelvin mode amplification keeping the initial rotational symmetry of the condensate densities.
Figure \ref{varicose} demonstrates the nonlinear development of the second component in the CFI induced only by the varicose mode amplification for $Q\xi=1.47$.
 The density modulation caused by the varicose mode amplification makes deep troughs in the density of the second component.
 The phase $\arg\Psi_2$ drastically changes around the density troughs,
 and then the winding number of the phase across the system along the core-flow vortex is decreased
 after an annihilation of a vortex-antivortex pair in the cross section of the phase distribution in \Fig{varicose}.
 This phenomenon is an analog of the phase slippage of a superflow in a narrow channel,
 where the winding number of the superflow along the channel decays after vortices pass across the channel.

\subsection{Crossover regime ($L>1$)}
For $L>1$, the amplification of splitting modes ($|l|=L$) with complex frequencies can have a dominant effect on the dynamic instability.
 In particular, as was mentioned above, the splitting modes survive even when the second component is absent in the limit $r\to r_0$.
 Because we are interested in the phenomena characteristic not of the multiquantized vortices but of helical shear flows,
 we demonstrate here the instability dynamics where the splitting mode amplification is subdominant in the helical vortex sheet regime.

\begin{figure}[htbp]
 \centering
 \includegraphics[width=1.0 \linewidth]{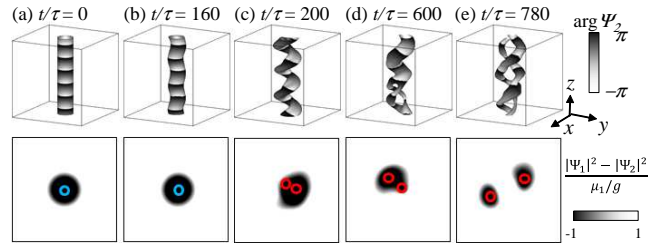}
 \caption
{
 (Color online) Development of the shear-flow instability for $L=2$ and $Q\xi=0.74$.
 Top panels show the density isosurface $|\Psi_1|=|\Psi_2|$ and the phase $\Theta_2=\arg\Psi_2$ of the second component on the isosurface.
 Bottom panels represent the distribution of the density difference $|\Psi_1|-|\Psi_2|$ around $r=0$ in the $z=0$ cross section.
 A circle in the cross section corresponds to the region containing vortices in the phase $\arg\Psi_1$ (not shown).
 (a) The cylindrical interface is straight in the initial state.
 (b) Amplification of the $|l|=1$ mode causes the helical deformation of the interface.
 (c) Vortex splitting is visible in the phase $\arg\Psi_1$ but not in the density difference.
 (d) The cylindrical interface partly splits into two singly quantized core-flow vortices in the place where the population of the second component becomes small locally.
 (e) Finally, the two core-flow vortices are twisted around each other.
 The system size is $40\xi$ in the $x$ and $y$ directions and  $51.2\xi$ in the $z$ direction.
}
 \label{L2dynamics}
\end{figure}

Figure \ref{L2dynamics} shows the dynamic instability developing from the helical shear flow for $L=2$ and $Q\xi=0.74$.
 The amplifications of three types of modes ($|l|=0,1,2$) are coexistent in the dynamics according to the plot of $L=2$ in \Fig{fig:diss_com}.
 A $|l|=1$ mode has the largest amplification rate in the state, and it induces a helical deformation of the cylindrical interface.
 In the early stage of the development,
 a deformation of the interface resulting from the amplification of the $|l|=1$ mode dominates over those of other modes [\Figs{L2dynamics}(b) and 10(c)].
 The radius of the cylindrical interface becomes thin locally as a result of density modulation caused by the varicose mode amplification.
 Then, vortex splitting occurs from the thin region [\Fig{L2dynamics}(d)].
 In the final stage [\Fig{L2dynamics}(e)],
 a double helix of core-flow vortices is formed,
 and the varicose motion is suppressed by the reduction of the relative velocity
 along the helical trajectory of the core, as is the case with the instability of the $L=1$ core-flow vortex in \Fig{kelvinVSvaricose}.

The time development of a $L=3$ core-flow vortex becomes more complex than that of the $L=2$ core-flow vortex.
 Although the process of the development depends on which modes are dominantly amplified in the first stage of the instability,
 we essentially obtained the same result in the final stage: A triple helix of $L=1$ core-flow vortices was formed.

\begin{figure}[htbp]
 \centering
 \includegraphics[width=1.0 \linewidth]{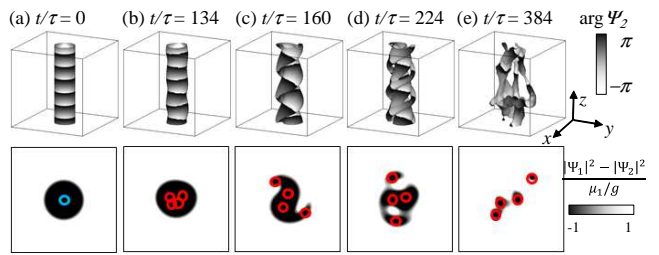}
\caption
{
 (Color online) Development of the instability from the helical vortex sheet for $L=4$ and $Q\xi=0.74$.
 (a)--(c) The straight cylindrical interface in the initial state is deformed into a unique three-dimensional structure as a result of the amplification of the $|l|=2$ mode.
 In the $z=0$ cross section, the interface forms a wavy pattern characteristic in quantum KHI.
 (d) Two single-quantized vortices are released outside from the tops of the waves,
 and then the nucleated vortices helically wind  the two vortices.
 (e) Finally,  a twisted bundle of four core-flow vortices appears with strong density modulation in the second component.
 The system is $76.8\xi$ in the $x$ and $y$ directions and, $51.2\xi$ in the the $z$ direction.
}
 \label{fig:dynamics_l4}
\end{figure}

The instability dynamics should change to those of quantum KHI when the interface radius $r_0$ increases with $L$.
 For our case,
 the signature of quantum KHI appears for $L=4$.
 Figure \ref{fig:dynamics_l4} shows the instability development of the helical shear flow for $L=4$ and $Q\xi=0.74$.
 In the early stage,
 a characteristic wavy pattern of quantum KHI \cite{2010TakeuchiPRB,2010SuzukiPRA} appears at the interface in the cross section of the cylindrical interface in \Fig{fig:dynamics_l4}(c).
 Since an $|l|=2$ ripple mode has the largest amplification rate,
 there appear two waves in the interface pattern
 and an characteristic twisted structure is formed in three dimensions.
 Two single-quantized vortices are nucleated outside from the tops of the waves [bottom of \Fig{fig:dynamics_l4}(d)],
 and then the nucleated vortices helically wind the two remaining vortices  around the $z$ axis [top of \Fig{fig:dynamics_l4}(d)].
 The remaining vortices form two single-quantized core-flow vortices
 and finally  a helically twisted bundle of four core-flow vortices appears [\Fig{fig:dynamics_l4}(e)].
 The population of the second component is not distributed equally among the four vortices
 and the density modulation is so strong that the second component sometimes forms a large droplet in the cores along some vortices.

\section{Conclusion and Discussion}
The helical shear flows can be classified into two states: a helical vortex sheet, where the radius of a cylindrical interface between the two components is large and the vorticity is localized at the interface, and
a core-flow vortex, where the population of the inside component flowing along the core becomes smaller and the vorticity is distributed around the core.
 These states develop into exotic instability phenomena in which various kinds of modes compete.

The linear stability of helical vortex sheets is well described with the effective theory of the helical KHI.
 When the superfluid velocity of the inside component exceeds a critical value,
 ripple modes trigger the dynamic instability, called the helical KHI.
 When the interface radius becomes comparable with the thickness of the interface,
 our numerical analysis revealed that the instability phenomena are affected strongly by the vortex-characteristic modes: Kelvin, splitting, and varicose modes.

 We showed the nonlinear development of the shear-flow instability by using three-dimensional numerical simulations.
 If the outside component has a single-quantized circulation in the core-flow vortex regime,
 the instability transforms the core-flow vortex from the initial straight line into a helix due to amplification of Kelvin modes.
 The direction of the helix is uniquely determined depending on the flow directions of the inside and outside components. 
 The helical deformation reduces the local relative velocity along the vortex core between the two components
 to suppress additional instabilities on the core-flow vortex.
 When the velocity of the second component becomes larger,
 varicose modes in addition to Kelvin modes are amplified in the instability development.
 The varicose modes induce density modulations, leading the  phase slippage in the second component
 to decrease the relative velocity more effectively than the helical deformation.
 These instabilities of the core-flow vortex caused by the amplification of Kelvin and/or varicose modes are called the CFI.
In the crossover regime between the helical KHI and the CFI,
 the primary difference between the KHI and the CFI is caused by the splitting modes,
 in which multiquantized vortices split into single-quantized vortices.
The vortex splitting effect  leads to a helically twisted bundle of single-quantized core-flow vortices in the final stage of the instability.

The helical shear-flow instability discussed in this paper is one of the fundamental hydrodynamic instabilities in multicomponent superfluid systems.
 This instability plays an important role in nonequilibrium dynamics of phase-separated two-component BECs containing quantized vortices
 when there is a larger population imbalance between the two components.
 For example, numerous core-flow vortices are nucleated via an annihilation of two interfaces (a domain wall and an antidomain wall) \cite{VortexformationJLTP, VortonPRA, TachyonPRL, TachyonJLTP}.
 In particular, the CFI must be crucial to understand the dynamics and stability of a vorton \cite{Metlitski2004JHEP,Bedaque2012JPB}, a loop of core-flow vortex.
These situations can be realized in two-component BECs with the experimental technique that was followed in Ref. \cite{B.P.Anderson2001PRL},
where the nodal plane in one component was filled with the other component and then
the filling component was selectively removed by using a resonant laser beam.
 Moreover, details of the instability developments can be investigated experimentally in cigar-shaped condensates,
in which a multiquantized vortex with large circulations is prepared in a component by using Laguerre-Gaussian beams or topological phase imprinting \cite{M.F.Andersen2006PRL, Leanhardt2002PRL, Mottonen2007PRL, Xu2008_2010PRA}.
 Relative motion along the vortex core can be realized, e.g., by employing a technique similar to that of \Ref{Hamner2011PRL} by utilizing the different Zeeman shifts between the two components.
 These realizations of the helical shear-flow instability provide an essential step prior to establishing modern quantum hydrodynamics \cite{phys_re} and lead to new physical insights not only into multicomponent BECs but also into other multicomponent superfluid systems.

\acknowledgments
This work was supported by the ``Topological Quantum Phenomena'' (No. 22103003) Grant-in Aid for Scientific Research on Innovative Areas from the Ministry of Education, Culture, Sports, Science and Technology (MEXT) of Japan.

\end{document}